\documentclass[journal,12pt]{IEEEtran}
\usepackage{bm}
\usepackage{cite} 
\usepackage{amsmath}
\usepackage{amssymb}
\usepackage{graphicx}
\usepackage{subfigure}
\usepackage{algorithmic}
\usepackage{graphicx}
\usepackage{graphics}
\usepackage{textcomp}
\usepackage{subfigure}
\usepackage{tabularx}
\usepackage{times}
\usepackage{multirow}
\usepackage{makecell}
\usepackage{bm}
\usepackage{latexsym}
\usepackage{booktabs}
\usepackage{threeparttable}
\usepackage{epsf}
\usepackage[ruled]{algorithm2e}
\usepackage{color,colortbl}
\usepackage{float}
\begin{document}
	%
	\title{Risk-Aware and Energy-Efficient AoI Optimization for Multi-Connectivity WNCS with  Short Packet Transmissions}
	%
	%
	%
	\author{
		Jie Cao,~\IEEEmembership{Member,~IEEE},
		Xu Zhu,~\IEEEmembership{Senior Member,~IEEE},\\
		Sumei Sun,~\IEEEmembership{Fellow,~IEEE}, Ernest Kurniawan,~\IEEEmembership{Senior Member,~IEEE},\\ Amnart Boonkajay,~\IEEEmembership{Member,~IEEE}
		\thanks{Jie Cao, Sumei Sun, Ernest Kurniawan and  Amnart Boonkajay are with the Institute of Infocomm Research, Agency for Science,
			Technology and Research, Singapore 138632.}\\
			\thanks{Xu Zhu 	is with the School of Electronic and Information Engineering, Harbin Institute of Technology, Shenzhen
		518055, China (Corresponding author: Xu Zhu, e-mail: xuzhu@ieee.org).}
	       }
		

	
	\maketitle
	\vspace{-20pt}
	\begin{abstract}

		Age of Information (AoI) has been proposed to quantify the freshness of information for 	emerging real-time applications such as remote monitoring and control in wireless networked control systems (WNCSs).
		Minimization of the average AoI and its outage  probability can ensure timely and stable transmission.
		Energy efficiency (EE)  also plays an important role in WNCSs, as many
		devices are featured by low cost and
		limited battery.  
		Multi-connectivity over multiple links enables a decrease in AoI, at the cost of energy. 
		We tackle the unresolved problem of selecting the optimal  number of connections {that is both AoI-optimal and energy-efficient}, while avoiding risky states.
		To address this issue, the average AoI and peak AoI (PAoI), as well as PAoI violation probability are formulated as functions of the number of connections.
		Then the EE-PAoI ratio is introduced to allow a tradeoff between AoI and energy, which is maximized by the proposed risk-aware, AoI-optimal and energy-efficient connectivity scheme.
		To obtain this, we analyze the property of the formulated EE-PAoI ratio and prove the  monotonicity of  PAoI violation probability. 
		Interestingly, we reveal that the  multi-connectivity scheme is not always preferable, and  the signal-to-noise ratio (SNR) threshold that  determines the selection of the multi-connectivity scheme is derived as a function of the coding rate. 
		Also, the optimal number of connections is obtained and  shown to be a decreasing function of the transmit power.
		Simulation results demonstrate that the proposed scheme enables more than 15 folds of EE-PAoI gain at the low SNR than the single-connectivity scheme.  

	\end{abstract}
	

	%
	\IEEEpeerreviewmaketitle

	\section{Introduction}

	\IEEEPARstart{I}{nformation}
	 freshness is a critical performance metric  for real-time applications such as remote sensing and  control in wireless networked control systems (WNCSs). The outdated sensing packets carrying stale information may result in inaccurate tracking and incorrect  operations\cite{traffic_5G}. 
	A new performance metric that incorporates  delay and packet loss, age of information (AoI), was proposed to quantify  information freshness\cite{howoften,book}. The authors in \cite{8845114} have indicated that the estimation error in WNCSs can be expressed as a non-decreasing function of AoI. 
	Hence, minimization of the expected value of AoI can ensure accurate and timely updating. But the average AoI fails to characterize extreme events in WNCSs, which may lead to risky states and hence system downtime\cite{risk_a}.
	Therefore, it is critical to consider both the average AoI and risk-related AoI performance for ensuring timely transmission and stable control  in WNCSs.
	
	Energy efficiency (EE)  also plays an important role in WNCSs, as a large number of communication
	devices  are typically featured by low cost and
	limited battery\cite{9141319}.  
	The multi-connectivity scheme is recognized as an effective approach to improve AoI performance, by using  multiple links over wireless communication\cite{multi1}.   One typical multi-connectivity scheme is transmission diversity, which facilitates the simultaneous transmission of the same data over multiple links. This scheme can improve	transmission reliability and thus reduce AoI, 
	but at the expense of higher energy consumption\cite{multi2,SHS1,SHS2}.
	However, the quantitative impacts of the multi-connectivity scheme on the average and extreme AoI performance remain unknown, thereby hindering  the preference for the number of connections in terms of risk, freshness and energy.

	\subsection{Related Work}

 
		\emph{AoI and risk analysis of WNCSs:} 
	In \cite{8845114}, remote estimation error was characterized as a non-decreasing function of AoI in the single-loop WNCS.
	To capture the worst case of AoI,  peak AoI (PAoI) was further 
	introduced in \cite{howoften}.
	In addition to the average AoI and PAoI, the  PAoI violation probability  was investigated  in \cite{9328793} and \cite{9324753} to characterize the extremely damaging event with a low probability of occurrence and thus avoid risky states in WNCSs.
	Specifically, the optimal generation rate  that induces the minimal violation probability was  also found in \cite{9324753}.
	Furthermore, the AoI tail distribution was characterized and controlled to improve the stability of vehicular networks in \cite{8937801}.
	The authors in \cite{risk_a} proposed a risk-aware transmission strategy to minimize the average AoI and the expected tail loss of the AoI. 
	However, only the single-connectivity WNCS was considered in the aforementioned work.

	\emph{AoI for  multi-connectivity 
		systems:} 
	Adopting the multi-connectivity scheme via multiple links is envisioned to reduce 
	the average AoI/PAoI by enhancing the transmission degrees of freedom\cite{multi1,multi2,SHS1,SHS2}. 
    Then the stochastic hybrid system method was employed  to analyze the average AoI of multiple queues\cite{SHS2,SHS1}. 
	Thereafter, \cite{multi1} demonstrated  that employing transmission diversity via multiple links can lead to a reduction in the AoI.
	In \cite{multi2}, the effect of transmission diversity on status age  was analyzed for the  resource-constrained  network, where $c$ represents the number of connections. \cite{multi2} also indicated that increasing the number of connections is beneficial for improving AoI performance, but at the expense of generating more outdated  packets,  resulting in wasted resources.  Therefore, this motivates thinking of the energy-efficient connectivity scheme for the AoI-oriented system.  Nevertheless, the aforementioned work on AoI of multi-connectivity systems were based on the  infinite block length (IBL) 
	regime assumption without considering block error probability (BLEP). Hence, these are
	not applicable to the real-time WNCSs with finite block length (FBL). 
	Furthermore, the risk analysis for the multi-connectivity WNCS from an AoI perspective has not been performed in the previous work. 
	Therefore, the impact of the variability of the AoI distribution on multi-connectivity WNCSs remains an open problem.

	

	\emph{  AoI in short-packet systems:} 
	In   practical real-time scenarios such as remote sensing in WNCSs, the status update  is transmitted using FBL, with payload size typically ranging from 20 to 250 bytes\cite{block0}. 
    Although the adoption of short packets is anticipated to reduce transmission delay and hence AoI, the lack of coding capability of short packets may lead to inevitable BLEP and compromise AoI performance\cite{block0}.
	The average AoI with FBL 	was investigated for the single-connectivity system
    in \cite{wangrui},  based on which the AoI minimization scheme was proposed under the energy constraint\cite{10146018}. 
	Thereafter, coding rate optimization was applied in the single-connectivity system\cite{9126228} and multi-sensor system\cite{9024516} for  AoI minimization.
	In \cite{8845098} and \cite{block2}, the PAoI 
	violation probability was analyzed to avoid risky states for the single-connectivity short-packet system over fading channels, where the delay was  also considered in \cite{8845098}.
	For reducing the average AoI, the  optimal updating policy and  coding rate  were respectively  derived for a single-connectivity system in \cite{9013381} and \cite{9484506}. 
	In \cite{Cj_aoi}, we studied the correlation between the average delay  and the 
	average AoI, as well as their joint optimization for the single-connectivity  short-packet system.
	However, the aforementioned work on AoI in the short-packet system was based on  a single-connectivity model\cite{wangrui,10146018,8845098,block2,9126228,9024516,9013381,9484506,Cj_aoi}.
	In our previous work, either only the average AoI of two queues was considered\cite{9997497}, or the energy efficiency of multi-connectivity systems was not addressed\cite{ICC_AoI_workshop}.

	\emph{AoI-Energy Tradeoff:} There is very few work in literature studying the AoI-energy tradeoff for the multi-connectivity WNCS with short packet transmissions.
	In \cite{8648195},  a practical truncated automatic repeat request (TARQ) policy
	was presented to enable the AoI-energy tradeoff. 
	Based on this, a dynamic programming approach and  a threshold-based stationary policy were proposed to minimize the weighted sum of AoI and  energy consumption  in \cite{9141319} and \cite{9524472}, respectively.
	Then, two improved TARQ schemes, namely TARQ with the selection combining and the maximum ratio combining (MRC) techniques, were  provided to achieve a better AoI-energy tradeoff in \cite{8761106}.
	In \cite{10008202}, the optimal sleep-wake scheduling policy was proposed to minimize the average AoI while saving energy. Furthermore, data compression was adopted in \cite{9951137} to  decrease the AoI and improve the energy efficiency (EE).
	To further improve the EE of information updates, a practical online upload scheduler was proposed in \cite{9483628} to minimize the  energy consumption with the  constraints of information freshness.
	In \cite{9465806},  the age and energy tradeoff was investigated for multicast systems, where the  AoI-EE ratio was introduced and is minimized by optimizing the block length. However, the AoI-energy tradeoff in the multi-connectivity WNCS remains unexplored, and  the optimal number of connections with the consideration of risk, freshness and EE is still unknown,   especially for short-packet systems.

	\subsection{Contributions}
	
	In the real-time WNCS, status information and control command with limited data size need to be updated in a timely and efficient manner, meanwhile ensuring system stability. 
	Inspired by the above mentioned open questions, a  multi-connectivity WNCS with short packets over Rayleigh fading channels is considered in this paper, aiming to answer the following
	questions: \textit{(1) What is the quantitative impact of the multi-connectivity scheme on the average and outage  performances of AoI? (2) Under what condition is a multi-connectivity scheme preferable to a single-connectivity scheme with respect to AoI and EE?  (3) What is the optimal number of connections that ensures both AoI and EE performance while avoiding risky states?} We summarize the main contributions  as follows.
	
	\begin{itemize}
		\item
		We analyze both the average and outage performances of AoI in the multi-connectivity short-packet WNCS, to ensure timely and stable transmission. The transmission diversity with the MRC technique is adopted over   fading channels. Nevertheless,  the existing work on AoI of multi-connectivity 
		systems \cite{multi1,multi2,SHS1,SHS2} focused on the average AoI only and assumed IBL 
		with zero BLEP, leading to different queueing processes and   AoI performance, compared to the FBL case.
		Furthermore, the previous work on AoI with FBL  was conducted for the 	single-connectivity system\cite{wangrui,8845098,block2,9126228,9024516,9013381,9484506,Cj_aoi}, with a lack of analysis of the impact of multiple connections.

		\item
		We explore the tradeoff between AoI and energy in  multi-connectivity short-packet WNCSs.
		Motivated by the unclear relationship between AoI and energy in the multi-connectivity WNCS, 
		the AoI performance and energy consumption  are analyzed in closed forms.
        The advantages of the multi-connectivity WNCS are demonstrated by an analytical comparison with the $K$-repetition  scheme.
	    To ensure the stability of WNCSs,	we  derive  the  violation probability of PAoI and reveal the relationship between the risky state and AoI. 	For this purpose, the probability density function (PDF)  and cumulative distribution function  (CDF) of PAoI are also analyzed with respect to PAoI threshold and the number of connections.
		However,  previous work on the multi-connectivity system have not conducted a comprehensive analysis of the characteristics  of AoI distribution
		\footnote{Note that this paper has fundamental differences from our previous
			work in [24] and [25].  In this manuscript, we focus on the AoI analysis and optimization for a risk-aware and energy-efficient multi-connectivity WNCS. While [24] provides an investigation of AoI for diversity and multiplexing with dual queues, and [25] presents the AoI analysis for diversity under more queues. The sepcific differences lie in that,  1) the integration of communication and control is taken into account for a closed-loop WNCS; 2) an intensive analysis is provided on the relationship between AoI, energy efficiency in communication and control risk is provided; 3) the optimal number of connections is analyzed with joint consideration of AoI, energy efficiency and control risk.}
		\cite{multi1,multi2,SHS1,SHS2}. 

		\item
		We unveil the condition that the multi-connectivity scheme is of low AoI and high EE  while avoiding risky states. We also  provide an explicit preference for the number of connections in terms of AoI and EE.
		Toward the risk-aware, AoI-optimal and energy-efficient connectivity scheme,  the EE-PAoI ratio that combines the average PAoI  and EE is introduced, and proved to be a concave linear fractional problem (CLFP) in relation to the number of connections. 
		Particularly, the signal-to-noise ratio (SNR) threshold for the multi-connectivity scheme selection is derived, which depends on the coding rate.
		Consequently, a SNR-aware  connection number selection scheme is provided to maximize the EE-PAoI ratio  with the constraints of total energy consumption and PAoI violation probability. The optimal number of connections is shown to be decreasing with respect to the transmit power.
		Simulation results demonstrate that the proposed  connectivity scheme is able to avoid  risky plant states. Meanwhile it enables more than 15 folds of EE-PAoI gain at the low SNR than the single connection case.
	\end{itemize}

	The remainder of this paper is organized as follows. In
	Section II, we introduce the considered WNCS model. We then analyze the plant state and AoI for the WNCS in Section III, based on which a risk-aware, AoI-optimal and energy-efficient multi-connectivity scheme is proposed in Section IV.  Section V demonstrates our conclusions with simulation results. Finally, Section VI concludes
	the paper.
	
	\section{System Model}
	In this section, we first introduce the considered multi-connectivity short-packet WNCS. Thereafter, we introduce the considered control model and illustrate the evolution of AoI and wireless transmission model.
	
	\subsection{System Description}

	We consider a multi-connectivity WNCS  with short packets, as illustrated in Fig. 1, in which the average AoI is very useful in reflecting the control performance. 
	On the other hand, the risk occurs when the PAoI is larger than a threshold, based on the assumption that a control application outage occurs as soon as the PAoI is violated\cite{9478879,8845098}.  
	To avoid risky states and ensure timely transmission, we adopt the multi-connectivity scheme to minimize the average AoI and guarantee that the PAoI violation probability is lower than the threshold.
 {For the multi-connectivity scheme with $K$ transmission diversity, same packets are  transmitted over multiple links simultaneously to improve transmission reliability and information freshness\cite{9478879}.}
	Specifically, a single source (or sensor)
	updates fresh information to the controller through  multiple wireless links.  After estimating the current state based on the received sensor updates, the controller computes and transmits the control command to the actuator immediately. 
	\begin{figure}[htbp]
		\centering
		{\includegraphics[height=3.5cm]{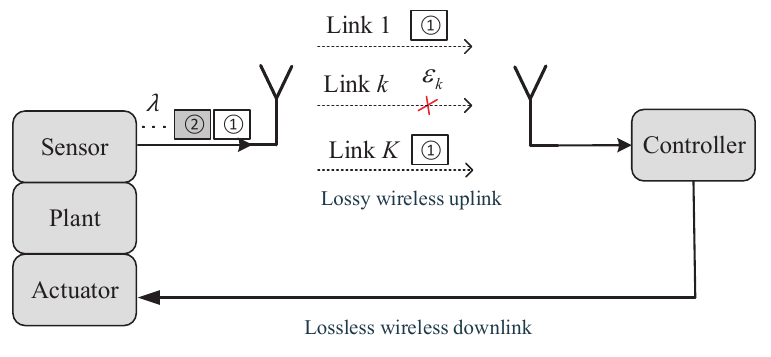}}
		\caption{Multi-connectivity enabled WNCS with short packets.}
		\label{fig_1}
	\end{figure}

	
   {The assumptions for the multi-connectivity short-packet WNCS are summarized as follows. 
	a) Packet generation  follows a Poisson process with  a mean arrival rate of 
	$\lambda$ \cite{error}. 
	b) Finite block length $m$ with modulation order $\Theta$ and $L$ bits information is considered \cite{block0}.  The inevitable BLEP  of the $k$-th link is denoted by $\epsilon_k$. 
	c) We concentrate on the uplink transmission in this paper,  \emph{i.e.,} packet losses may only occur between the sensor to the controller, following \cite{9305697}.
	The erroneous downlink channel will be incorporated in our future work.
	d) The non-preemptive Last-Come-First-Served (LCFS) scheme is adopted to reduce information age and avoid unbounded AoI under high arrival rates \cite{LCFS1}. Assuming a buffer size of zero,  newly arriving packets are	discarded when multiple links are busy.                 
	e) Multiple independent links without retransmission can be achieved by  orthogonal channels, and the MRC technique is adopted at  the controller to improve  transmission reliability\cite{8761106,7056486}.
	f) In order to reveal the impact of the number of connections on EE and AoI, synchronous transmission with equal power for each link is assumed, following \cite{9478879}.  A more practical system with tunable transmit power will be considered in the future.}
	
	\subsection{Control System Model}
	We consider the physical process modeled by the following time-slotted linear time-invariant (LTI) system\cite{8845114,9305697}
	\begin{equation}	\mathbf{x}_{n+1}=\mathbf{A}\mathbf{x}_n+\mathbf{B}\textbf{u}_{n}+\mathbf{w}_n,
	\end{equation}
	where $n$ is the  time index.
	$\mathbf{A}\in\mathbb{R}^{d\times d}$ and $\mathbf{B}\in\mathbb{R}^{d\times q}$  represent the system matrix and control input matrix, respectively.
	Besides, $\mathbf{x}\in\mathbb{R}^{d\times1}$ is the plant state with the dimension of $d$
	and 	$\mathbf{u}\in\mathbb{R}^{q\times1}$ denotes the control input with the dimension of $q$. $\mathbf{w}\in\mathbb{R}^{d\times q}$ is the normally distributed noise with the distribution $\mathcal{N}(0,\mathbf{R}_w)$, where $\mathbf{R}_w$ is the noise variance.
	Generally, $\mathbf{x}$  represents the deviation between the true value and the set value of the considered process, which is expected to be kept close to $\mathbf{0}$. Hence, the control outage (\emph{i.e.}, risky state) occurs when the  state is larger than a threshold. 
	In this paper, we consider one-step control, where the  control policy can be expressed  as $\mathbf{u}_n=\mathbf{G}_n\hat{\textbf{x}}_n$, where $\mathbf{G}_n=-\mathbf{A}/\mathbf{B}$ is the optimal control gain\cite{8865111}.
	Since the control performance is affected by the freshness of the received packets\cite{8845114}, we analyze the AoI model in the next subsection.

	\subsection{AoI Model}
\begin{figure}[htbp]
		\centering
		{\includegraphics[height=5cm]{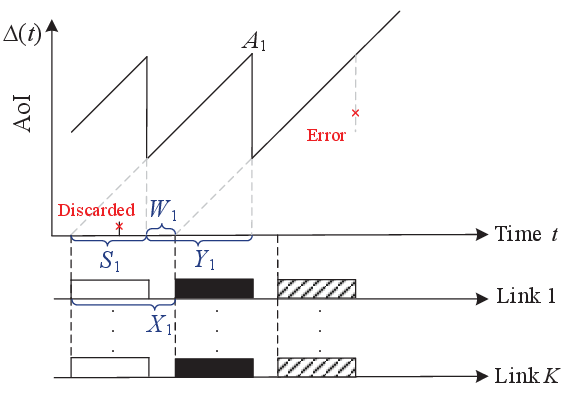}}
		\caption{The evolution of AoI and queueing process in the multi-connectivity WNCS with transmission diversity.
			The red `x' symbol on the left-hand side
		represents that the packet was dropped  due to  multiple  {links} being 
		busy. The right red `x' symbol on the right-hand side
		represents that the packet transmission  failed.
		}\label{fig_2_trans}
	\end{figure}

	The instantaneous AoI at the controller in the multi-connectivity WNCS is illustrated in Fig.  \ref{fig_2_trans}. With the assumption of a continuous-time state-update system, time $t\in[0,\infty)$ is allowed to take any positive value.
	Mathematically, instantaneous AoI can be expressed as $\Delta(t) = t - u(t)$, which represents  the duration since the generation time, $u(t)$, of the most recent successfully received packet.
	As shown in Fig. 2, the service time is denoted by $S_i$,  which represents the time interval between the generation and successful reception of the $i$-th update.  The inter-arrival time between two arrival packets is given by $X_i$. Also,  the 	inter-departure   time, $Y_i$, defined as the time interval  between two consecutive successful packet receptions.

		As shown in Fig. 2, the  PAoI of the $i$-th update is given by 
	$A_i=S_{i}+Y_i$.
	Hence, the time average PAoI $\bar{A}$ over the  time period of $\mathcal{T}$ is calculated as
	\begin{equation}
		\begin{aligned}
			\bar{A}= \mathbb{E}[Y]+\mathbb{E}[S].
		\end{aligned}\label{eq_3}
	\end{equation}


	The time-average AoI can be calculated by a sum of the 
	concatenation of trapezoids area $J_i$ over a time period $\mathcal{T}$. The $j$-th shadowed trapezoid area is calculated as
	$J_j=\frac{(Y_j+S_j)^2}{2}-\frac{S_{j+1}^2}{2}=\frac{Y_j^{2}}{2}+Y_j S_j$,
	where  $\{S_1,...S_j\}$ and $\{Y_1,...Y_j\}$ are independent and identically 
	distributed\cite{book}.  Then, the average AoI $\bar{\Delta}=\frac{1}{\mathcal{T}} \int_{0}^{\mathcal{T}} 
	\Delta(t)dt$ can be further expressed as
	\begin{small}
		\begin{equation}
		\label{eq_1}
		\begin{aligned}
		\bar{\Delta}=\lim _{\mathcal{T} \rightarrow 
			\infty}\frac{N(\mathcal{T})}{\mathcal{T}} \frac{1}{N(\mathcal{T})} 
		\sum_{i=1}^{N(\mathcal{T})} J_{j}=\frac{\mathbb{E}[Y^{2}]}{2 \mathbb{E}[Y]}+\mathbb{E}[S],
	\end{aligned}
\end{equation}
	\end{small}where $\mathbb{E}[\cdot]$ denotes the expectation operation and  $N(\mathcal{T})$ represents the number of successfully  	received informative packets within $\mathcal{T}$.   

%


\subsection{Wireless Transmission Model}

Recall that we assume that the uplink (sensor-controller) is imperfect in the WNCS, where both the transmission delay and error are considered.
To enhance the transmission reliability,  multiple independent transmissions via orthogonal links are employed with MRC technique at the receiver.
For the sake of analysis, the normal approximation of BLEP with FBL  is adopted\cite{9126228}, which  is  approximated by converting the Q-function into the segmented linear function. With given coding rate $R=L/m$ and received SNR  $\gamma$, we have
 \begin{equation}\label{eq_error}
	\epsilon(\gamma)
	\approx\left\{
	\begin{aligned}
		&	1, &\gamma<\varpi+1/(2\phi)\\
		&	\phi(\gamma-\varpi)+\frac{1}{2}, &\varpi+\frac{1}{2\phi}\le\gamma<\varpi-\frac{1}{2\phi}\\
		&	0, &\gamma>\varpi-1/(2\phi),
	\end{aligned}
	\right.
\end{equation}
where $\varpi=e^{R}-1$,  $\phi=-\sqrt{\frac{m}{2\pi(e^{2R}-1)}}$.

Considering the  total transmit power $P_\textrm{total}$ and average noise variance $\sigma^2$,  the average SNR of each connection is denoted by $\bar{\gamma}=\frac{P_\textrm{total}}{K\sigma^2}=\frac{P_\textrm{t}}{\sigma^2}$ with equal power allocation among  $K$  {connections}. Here, $P_\textrm{t}$ denotes the transmit power of each link. 
Therefore, the instantaneous SNR of each  {connection}, \emph{i.e.}, $\gamma=\bar{\gamma}\left|h\right|^2$,  follows an exponential distribution,  where $h$ represents the instantaneous channel gain following a complex Gaussian distribution with  zero mean and unit variance. 
With $K$ orthogonal channels, the received SNR follows the   Erlang distribution\cite{book}, \emph{i.e.}, $f_K(\gamma)=\frac{\gamma^{K-1}}{{\bar{\gamma}}^K(K-1)!}e^{-\frac{\gamma}{\bar{\gamma}}}$.
Then, the average BLEP $\bar{\epsilon}_K$ of the short-packet WNCS with $K$ connections is calculated by 
$	\bar{\epsilon}_K=\int_{0}^{\infty}f_K(\gamma)\epsilon(\gamma)d\gamma$, which is derived in detail in \cite{ICC_AoI_workshop} and can be further expressed as 
\begin{small}
\begin{equation}
	\begin{aligned}
		\bar{\epsilon}_K&=\frac{1}{(K-1)!}\int_{0}^{\infty}\frac{\gamma^{K-1}}{{\bar{\gamma}}^K}e^{-\frac{\gamma}{\bar{\gamma}}}\epsilon(\gamma)d\gamma\\
		&\approx1-\frac{	\Gamma\left(K,-\ln(\omega_1)\right)+(0.5-\varpi\phi+K\phi\bar{\gamma})\Sigma+	\phi\bar{\gamma}\Xi}{(K-1)!},\label{eq_6}
	\end{aligned}
\end{equation}
\end{small}where $\Sigma=	\Gamma\left(K,-\ln(\omega_2)\right)-\Gamma\left(K,-\ln(\omega_1)\right)$ and $\Xi=(-\ln(\omega_2))^K\omega_2-(-\ln(\omega_1))^K\omega_1$ with $\omega_1=e^{-\frac{1}{\bar{\gamma}}(\varpi+\frac{1}{2\phi})}$ and
$\omega_2=e^{-\frac{1}{\bar{\gamma}}(\varpi-\frac{1}{2\phi})}$. Also, $\Gamma(K,-\ln(\omega_1))=\int_{-\ln(\omega_1)}^{\infty}x^{K-1}e^{-x}dx$   represents the incomplete gamma function.
For the single-connectivity case, the average BLEP in (5) is reduced to (6), given by \cite{9126228}
\begin{equation}
	\begin{aligned}
		\bar{\epsilon}&=\int_{0}^{\infty}\frac{1}{\bar{\gamma}}e^{-\frac{\gamma}{\bar{\gamma}}}\epsilon(\gamma)d\gamma
		\approx 1+\phi\bar{\gamma}(\omega_1-\omega_2).
	\end{aligned}\label{E_error}
\end{equation}

						\section{State and AoI Analysis  for Multi-Connectivity Enabled WNCSs}
						
						In order to avoid risky states in the WNCS and ensure its stability, we first analyze the evolution of plant state and reveal the relationship between  plant state and AoI. 
						Then the  distribution of AoI for the multi-connectivity WNCS is analyzed with respect to the number of connections. 

						\subsection{Risky State Analysis}
						
						For the stability of WNCS, we aim to maintain the plant state close to zero (\emph{e.g.}, frequency deviation in smart grid). 
						On one hand, the time-average plant state is required to be minimized to reduce the difference between the true state and preset value, \emph{i.e.},  $\underset{n\rightarrow\infty}{\lim}\frac{1}{n}\sum_{j}^{n-1}||\mathbf{x}_j||_2^2$\cite{8845114}. On the other hand, extreme plant states that exceed thresholds should also be avoided.
						
						Due to the uplink transmission error and delay, the controller may not be able to receive the sensor's updates  timely and accurately, failing to track the plant state in real-time and calculate the correct control commands. To  tackle this issue, we consider a controller with an integrated estimator, which can estimate the current plant state by leveraging the previously received packets.
						At the controller
						side, we analyze the AoI at discrete time instant $n$, denoted by $\Delta_n=n-\tau_n, n=1,2,\cdots,N$, where $\tau_n=g_{\underset{i}{\max}\left\{d_i\leq n\right\}}$ is the generation time of the most recently received packet at the controller before $n$.
						Then we extend the conclusion in \cite{8845114}         and provide Proposition 1 to illustrate the relationship between plant state and AoI.

						\textbf{Proposition 1}   The  plant state of the WNCS can be expressed as a non-decreasing function of sensor's AoI.
						The plant-state covariance is given by
						\begin{equation}
							\mathbb{E}[\mathbf{x}_n\mathbf{x}^{\top}_n]=\sum_{i=1}^{\Delta_n}\mathbf{A}^{i}\mathbf{R}_w(\mathbf{A}^{\top})^i+\mathbf{R}_w,
						\end{equation}
						where $\Delta_n$ denotes the AoI of the received sensor's packet at the controller before $n$.
						
						\textit{Proof}: Based on the received packets, the controller estimates the current plant state as $
						\hat{\mathbf{x}}_{n}=\mathbf{A}^{\Delta_{n}}\mathbf{x}_{n-\Delta_{n}}+\mathbf{B}\mathbf{u}_{n-1}+\cdots+\mathbf{A}^{\Delta_{n-1}}\mathbf{B}\mathbf{u}_{n-\Delta_n}$.
						Based on (1), the dynamic of plant state can be expressed as
						${\mathbf{x}}_{n}=\mathbf{A}^{\Delta_{n}}\mathbf{x}_{n-\Delta_{n}}+\sum_{j=1}^{\Delta_n}\mathbf{A}^{j-1}\mathbf{w}_{n-j}+\mathbf{B}\mathbf{u}_{n-1}+\cdots+\mathbf{A}^{\Delta_{n-1}}\mathbf{B}\mathbf{u}_{n-\Delta_n}$.
						Then the estimation error at the controller is given by
						$  \mathbf{e}_n=\mathbf{x}_n-\hat{\mathbf{x}}_n=\sum_{i=1}^{\Delta_n}\mathbf{A}^{i-1}\mathbf{w}_{n-i}$, which results in $
						\mathbb{E}[\mathbf{e}_n\mathbf{e}^{T}_n]=\sum_{i=1}^{\Delta_n}\mathbf{A}^{i-1}\mathbf{R}_w(\mathbf{A}^{T})^{i-1}$.
						By feeding back computed control signal $\mathbf{u}_n=-\mathbf{A}/\mathbf{B}\hat{\textbf{x}}_n$ to the actuator, we can then
						express the closed-loop dynamics as $ \mathbf{x}_{n+1}=\mathbf{A}(\mathbf{x}_n-\hat{\mathbf{x}}_n)+\mathbf{w}_n=\mathbf{A}\mathbf{e}_n+\mathbf{w}_n$, which yields the covariance of plant state in (7). $\hfill\blacksquare$

						Proposition 1 implies that minimizing the time-average plant state  is equivalent to minimizing the average AoI. Hence,  the plant state should be lower than a threshold to avoid risky states. To ensure this, the PAoI should also be lower than a preset threshold, which is related to the PAoI violation probability.
						Hence, in the following subsections, we focus on the average AoI/PAoI and PAoI violation probability at the controller in the multi-connectivity WNCS for timely transmission and stable control.
						
						\subsection{Average AoI}

						We focus on the average AoI at the controller in the multi-connectivity WNCS.
						As shown in Fig. \ref{fig_2_trans}, the queueing process in the multi-connectivity scheme with synchronous transmission is similar to the single-connectivity case\cite{wangrui,9478879}.
						 Hence, 
						the  multi-connectivity WNCS can be viewed as an $M/D/c$ model.
					    In this subsection, we derive the average AoI/PAoI for the multi-connectivity WNCS without retransmission. Also, the analysis of the single-connectivity scheme with ARQ retransmission and $K$-repetition are provided for comparison.
						
					\subsubsection{Multi-connectivity WNCS without retransmission}	
					With the non-retransmission (NR) policy, {the service time is deterministic and represented as  $S^{\text{NR}}=M=mT_\textrm{s}$},  with $T_\textrm{s}$ 	denoting the symbol 	duration. The inter-departure time is denoted by $\mathbb{E}[Y^{\text{NR}}]=\frac{\frac{1}{\lambda}+M}{1-\bar{\epsilon}_K}$, with the average BLEP  of the   multi-connectivity  WNCS $\bar{\epsilon}_K$. 
					Based on the analysis of the single-connectivity scheme without retransmission in \cite{wangrui},  we have Theorem 1.
						
						\textbf{Theorem 1}
						The  average PAoI and AoI of the multi-connectivity short-packet  WNCS over $K$   links without retransmission  are respectively given by
						\begin{equation}
							\begin{aligned}
								\bar{A}^{\text{NR}}_{c=K}&=\frac{\frac{1}{\lambda}+M}{1-\bar{\epsilon}_K}+M,
							\end{aligned}\label{eq_7}
						\end{equation}
					and
						\begin{equation}
						\begin{aligned}
							\bar{\Delta}^{\text{NR}}_{c=K}&=\frac{(1+\lambda M)(1+\bar{\epsilon}_K)}{2\lambda(1-\bar{\epsilon}_K)}+\frac{1+2\lambda M+2\lambda^2 M^2}{2\lambda+2\lambda^2 M}.
						\end{aligned}\label{eq_5}
					\end{equation}
						
						\textit{Proof:} Please refer to Appendix A.$\hfill\blacksquare$

							 Theorem 1 indicates that the average AoI/PAoI depends on the number of connections, based on which
							 their properties are concluded in Corollary 1.

						\textbf{Corollary 1}
						{The average AoI/PAoI for the multi-connectivity WNCS without retransmission is monotonically decreasing with respect to the number of connections. With $K\in\left[1,\infty\right)$ connections, the range of the average AoI/PAoI is given by
							
							\renewcommand\arraystretch{1.6}
							\begin{table}[h]
								\centering
								\caption{Range of the average AoI/PAoI in multi-connectivity WNCS without retransmission}
								\begin{tabular}{|l|l|l|}
									\hline
									& Average AoI & Average PAoI \\ \hline
									$K=1$                         & $\frac{(1+\lambda M)(1+\bar{\epsilon})}{2\lambda(1-\bar{\epsilon})}+\frac{1+2\lambda M+2\lambda^2 M^2}{2\lambda+2\lambda^2 M}$   & $\frac{\frac{1}{\lambda}+M}{1-\bar{\epsilon}}+M$    \\ \hline
									$K\rightarrow\infty$ & $\frac{1+\lambda M}{2\lambda}+\frac{1+2\lambda M+2\lambda^2 M^2}{2\lambda+2\lambda^2 M}$   &  $\frac{1}{\lambda}+2M$    \\ \hline
								\end{tabular}
							\end{table}
						
							\textit{Proof:}  
						The first-order derivative of the average PAoI in relation to the average BLEP is calculated as $\frac{\partial	\bar{A}^{\text{NR}}_{c=K}}{\partial\bar{\epsilon}_K}=\frac{\frac{1}{\lambda}+M}{(1-\bar{\epsilon}_K)^2}>0$.
						The monotonicity of the average PAoI with respect to the number of connections is guaranteed since a larger number of connections results in a higher received signal-to-noise ratio and transmission reliability.
						A similar approach can be applied to the average AoI case.
						Then, substituting $\bar{\epsilon}_K=0$ and (6) into (8) and (9) yields the extreme values in Table I.  $\hfill\blacksquare$
						
						Corollary 1 indicates that the  lower bounds of the average AoI/PAoI are constrained  by the service time $M$ and arrival rate $\lambda$, implying that it is not always useful to reduce the information age by increasing the number of connections.
						
						\subsubsection{Single-connectivity WNCS with ARQ retransmission}
						For the case of ARQ retransmission, the source keeps transmitting packets untill the packet is successfully received, which requires the immediate feedback\cite{8648195}. The average AoI and PAoI for the single-connectivity WNCS with ARQ retransmission is given in Thereom 2.
						
							\textbf{Theorem 2}
						The  average PAoI and AoI of the single-connectivity   WNCS with ARQ retransmission over fading channels  are respectively given by
							\begin{equation}
							\begin{aligned}
								\bar{A}^{\text{ARQ}}&=\frac{2M}{1-\bar{\epsilon}}+\frac{1}{\lambda},
							\end{aligned}
						\end{equation}
					and
						\begin{small}
						\begin{equation}
							\begin{aligned}
								\bar{\Delta}^{\text{ARQ}}&=\frac{2(1-\bar{\epsilon})+\lambda^2M^2+2\lambda M(1-\bar{\epsilon})}{2\lambda(1-\bar{\epsilon}+\lambda M)}+\frac{M(1+\bar{\epsilon})}{1-\bar{\epsilon}}.
							\end{aligned}
						\end{equation}
						\end{small}
										
						\textit{Proof:} Please refer to Appendix B.$\hfill\blacksquare$
						
					\subsubsection{Single-connectivity WNCS with $K$-Repetition}
			To avoid feedback overhead and enhance reliability, $K$-repetition (KR) is  considered.
			 With $K$ repetitions, the service time is given by $S^{\text{KR}}=KM$, and the inter-departure time is provided as $\mathbb{E}[Y^{\text{KR}}]=\frac{\frac{1}{\lambda}+KM}{1-\bar{\epsilon}^K}$. 
			Then the average AoI and PAoI are provided in Theorem 3.
			
				\textbf{Theorem 3}
			The  average PAoI and AoI of the single-connectivity WNCS with $K$ repetitions over fading channels  are respectively given by
		 \begin{equation}
				\begin{aligned}
					\bar{A}^{\text{KR}}&=\frac{\frac{1}{\lambda}+KM}{1-\bar{\epsilon}^K}+KM,
				\end{aligned}
			\end{equation}
		and
		\begin{small}
			\begin{equation}
				\begin{aligned}
					\bar{\Delta}^{\text{KR}}&=\frac{(1+\lambda KM)(1+\bar{\epsilon}^K)}{2\lambda(1-\bar{\epsilon}^K)}+\frac{1+2\lambda K M+2\lambda^2 K^2M^2}{2\lambda+2\lambda^2 KM}.
				\end{aligned}
			\end{equation}
		\end{small}
			\textit{Proof:}   Substituting $S^{\text{KR}}$ and $\mathbb{E}[Y^{\text{KR}}]$ into (2) and (3) yields the average PAoI in (12) and AoI (13). 	$\hfill\blacksquare$
%

	
	\subsubsection{Comparison}						
Then we compare the above three approaches in terms of the average PAoI, as  it exhibits a similar trend with the average AoI but more tractable\cite{9997497}.	
Based on Theorems 1, 2 and 3, we have Corollaries 2 and 3. 

	\textbf{Corollary 2}
The single-connectivity WNCS with ARQ retransmission achieves a lower average PAoI but induces more information exchange overhead compared with the non-retransmission case.

	\textit{Proof:} Based on (8) and (10), the gap is given by $G_{A-N}=\frac{M-\frac{1}{\lambda}}{1-\bar{\epsilon}}+\frac{1}{\lambda}-M$.
	Since $\frac{1}{\lambda}\geq M$ is held for queueing stability, we have $G_{A-N}<0$ with a non-zero $\bar{\epsilon}$.
	$\hfill\blacksquare$

	\textbf{Corollary 3}
The multi-connectivity WNCS achieves a lower average PAoI than the single-connectivity WNCS with $K$-repetitions.

\textit{Proof:} For the $K$-repetition scheme, more repetitions improve transmission reliability but induce larger delay. Hence, we mainly compare the case of $K=2$, as the gain from repetition is more limited in other cases. According to (5), the average BLEP with $K=2$ is given by $\bar{\epsilon}_{2}=1+\phi(2\bar{\gamma}+\varpi)(\omega_1-\omega_2)+\frac{\omega_1+\omega_2}{2}$. Then $\bar{\epsilon}_{2}<\bar{\epsilon}^2$ can be obtained based on (6), which yields   $\bar{A}^{\text{NR}}_{c=2}<	\bar{A}^{\text{KR}}$. Then, the other cases can be proved in a similar manner.
$\hfill\blacksquare$

Corollary 3 demonstrates the advantage of the multi-connectivity scheme.
To reduce the information exchange overhead\cite{9126228}, we focus on the multi-connectivity WNCS without retransmission  and omit the superscripts hereafter. ARQ retransmission will be incorporated in our future work.


									\subsection{{PAoI Violation Probability}}
									
									PAoI violation probability is utilized to measure the outage performance of AoI, which can be limited to avoid risky states in WNCSs.  By calculating the probability that PAoI exceeds a preset threshold $\zeta$, PAoI violation probability is represented as ${\rm{Pr}}_{c=K}[A>\zeta] = \lim_{i\rightarrow\infty} {\rm{Pr}}[A_i>\zeta]$.  $A_i$
									represents the
									instantaneous PAoI of the $i$-th update. Based on the queueing process and packet assignment of the multi-connectivity WNCS, we have Theorem 4.

									\textbf{Theroem 4}
									{The PDF and CDF of PAoI  for the multi-connectivity WNCS with $K$   links  are respectively given by
										\begin{equation}
											f(x)_{c=K}=
											\left\{
											\begin{aligned}
													&\lambda(1-\bar{\epsilon}_K)e^{\lambda(1-\bar{\epsilon}_K)(2M-x)}, &x>2M,\\
												&	0, &x\leq2M,
												\end{aligned}\label{eq_12}
											\right.
										\end{equation}
										and  
														\begin{equation}
												F(x)_{c=K}=
											\left\{
											\begin{aligned}
													&1-e^{\lambda(1-\bar{\epsilon}_K)(2M-x)}, &x>2M,\\
												&	0, &x\leq2M,
												\end{aligned}\label{eq_13}
											\right.
										\end{equation}
								Then the PAoI violation probability   is formulated as an exponential function of threshold $\zeta$ ($\zeta>2M$), given by
										\begin{equation}
											{\rm{Pr}}[A>\zeta]_{c=K}=e^{\lambda(1-\bar{\epsilon}_K)(2M-\zeta)}.\label{eq_14}
										\end{equation}
									}
								
									\textit{Proof:} From Theorem 1, it is deduced that PAoI is determined by the service time $S^{\text{NR}}=M$ and inter-departure time $Y^{\text{NR}}$, which consists of the waiting
									time $W^{\text{NR}}$ and service time $S^{\text{NR}}$. Therefore, the instantaneous PAoI is greater  than twice the service
									time, represented as $A> 2M$, and the waiting time can be expressed as $W^{\text{NR}}=A-2M$. Due to the memoryless property of the Poisson process, the waiting time follows an
									exponential distribution with a rate of $\lambda$ under successful transmission conditions.
									 The PDF of PAoI is calculated accordingly, as shown in (14). Then calculating the integral of the PDF  yields the CDF in (\ref{eq_13})  and violation probability in (\ref{eq_14}).
									$\hfill\blacksquare$ 
									
									%
									Theorem 4 indicates that a larger PAoI threshold and a lower BLEP result in a lower PAoI violation probability.      
									Then, its property with respect to the number of connections is analyzed in Corollary 4.   
									
									\textbf{Corollary 4}
									{The PAoI violation probability is decreasing in relation to the number of connections. With $K\in\left[1,\infty\right)$ connections, the probability of PAoI violation ranges from $e^{\lambda(2M-\zeta)}$ to $e^{\lambda\phi\bar{\gamma}(\omega_2-\omega_1)(2M-\zeta)}$.
										
										\textit{Proof:} The first-order derivative of the PAoI violation probability in relation to the average BLEP is provided as $\frac{\partial	{\rm{Pr}}[A>\zeta]_{c=K}}{\partial\bar{\epsilon}_K}=\lambda(\zeta-2M)e^{\lambda(1-\bar{\epsilon}_K)(2M-\zeta)}>0$, with $\zeta>2M$.
										Thanks to the monotonicity of the average BLEP with respect to the number of connections, 
										we can obtain the infimum PAoI violation probability by substituting $\bar{\epsilon}_K=0$ into equation (\ref{eq_14}). Similarly, by substituting the average BLEP from equation (7) into equation (\ref{eq_14}), we can obtain the supremum PAoI violation probability.
										$\hfill\blacksquare$
										
								
								

								\section{Risk-Aware, Timely and Efficient Connectivity Scheme}

								Based on the analysis of risky state and AoI distribution  in Section III, it is concluded that increasing the number of connections is beneficial in reducing the average PAoI/AoI, and the PAoI violation probability, at the cost of energy. Also, it is impractical to 
								implement infinite number of connections in realistic scenarios, which also impairs the EE. Therefore, in this section, we aim to find the optimal number of connections that achieves the best AoI performance and EE while avoiding risky states.
								To obtain this, the EE-PAoI ratio is introduced\cite{9465806}, based on which the optimization problem is formulated, and its properties are analyzed. Then the risk-aware connectivity scheme that attends to both AoI and EE is proposed.

								\subsection{ Problem Formulation}
								We first introduce EE to evaluate the transmission efficiency of the multi-connectivity scheme, 		which is defined as the ratio of effective transmission rate and the total power consumption, given by $\frac{L(1-\bar{\epsilon}_K)}{MP_\textrm{total}}$. 				As stated in subsection II-C, equal power allocation between multiple links is assumed.  Hence, the total transmit power $P_{\textrm{total}}$ is the sum of the transmit power $P_\textrm{t}$ of $K$ connections, given by $P_{\textrm{total}}=KP_\textrm{t}$, where the cost of the deployment of multiple connections is neglected\cite{multi2}. 
								To facilitate the analysis of the multi-connectivity scheme, we
								focus on PAoI performance in this section, while the average AoI  is closely related to the average PAoI and therefore is not analyzed here\cite{7282742}.
								Therefore, we introduce the EE-PAoI ratio to improve the AoI-energy tradeoff, given by $\eta=\frac{L(1-\bar{\epsilon}_K)}{MK\bar{A}_KP_\textrm{t}}$, following \cite{9465806}.
								In order to optimize the average PAoI and EE simultaneously while avoiding risky states, the problem can be formulated as
								\begin{flalign} 
									\;\;\;\;	\mathcal{P} 1:
									\;\;
									&\underset{{K}}{\mathop{\max}}\;\;\;\;\;\eta=\frac{L(1-\bar{\epsilon}_K)^2}{MKP_\textrm{t}({\frac{1}{\lambda}+M}+M(1-\bar{\epsilon}_K))}&
									\label{eq_19}
								\end{flalign}
								\[
								\begin{split}
									\text{s.t.}\;\;\;\;\;& 
									(C1):	KP_\textrm{t}\leq P_{\max},\\
									&(C2):	{\rm{Pr}}[A>\zeta]\leq \rm{Pr}_{\max},\\
									&(C3):  K\in\mathbb{N}_+,\\
								\end{split}
								\] 
								where ($C1$) and ($C2$) constraint the maximum allowable total transmit power $P_{\max}$ and PAoI violation probability $\rm{Pr}_{\max}$, respectively. ($C3$) addresses that the number of connections belongs to the positive integer set $\mathbb{N}_+$.
								
								Based on (17), it is concluded that the energy-efficient connectivity scheme for AoI-oriented systems is obtained by maximizing the EE-PAoI ratio. A higher EE-PAoI ratio indicates the better ability to maintain the low average PAoI and high EE. Based on this, the condition that the multi-connectivity scheme is AoI-optimal and energy-efficient is analyzed, as shown in Proposition 2.
								
								\textbf{Proposition 2} 
								In the case of low average SNR, the multi-connectivity scheme with MRC is {more energy efficient} in AoI-oriented systems.
								As average SNR $\bar{\gamma}$ increases to $\bar{\gamma}_\textrm{thr}=2e^{R}-2$, the single-connectivity scheme is preferable in terms of the EE-PAoI ratio.

								\textit{Proof:} {First, we introduce the EE-PAoI gain of the multi-connectivity scheme compared to the single-connectivity scheme, denoted by} $G_\eta=\frac{(\frac{1}{\lambda}+M)(\frac{1-\bar{\epsilon}_K}{1-\bar{\epsilon}})+M(1-\bar{\epsilon}_K)}{K(\frac{1}{\lambda}+M)(\frac{1-\bar{\epsilon}}{1-\bar{\epsilon}_K})+KM(1-\bar{\epsilon})}$. 
								Particularly, we have $G_\eta\approx\frac{ (1-\bar{\epsilon}_K)^2(2-\bar{\epsilon})}{K(1-\bar{\epsilon})^2(2-\bar{\epsilon}_K)}$ at high arrival rate, and  $G_\eta>1$ indicates that the multi-connectivity scheme is better in terms of PAoI and EE.
								Without loss of generality, we prove $(1-\bar{\epsilon}_K)^2(2-\bar{\epsilon})>K(1-\bar{\epsilon})^2(2-\bar{\epsilon}_K)$ in the following, which is the most difficult to achieve when $K$ equals 2. In the case of $K=2$, we have $\bar{\epsilon}=1+\phi\bar{\gamma}(\omega_1-\omega_2)$ and $\bar{\epsilon_2}=1+\phi(2\bar{\gamma}+\varpi)(\omega_1-\omega_2)+0.5(\omega_1+\omega_2)$. 
								Then we unveil the condition that $(\phi(2\bar{\gamma}+\varpi)(\omega_1-\omega_2)+0.5(\omega_1+\omega_2))^2(1-\phi\bar{\gamma}(\omega_1-\omega_2))>2\phi^2{\bar{\gamma}}^2(\omega_1-\omega_2)^2(1-\phi(2\bar{\gamma}+\varpi)(\omega_1-\omega_2)-0.5(\omega_1+\omega_2))$ based on the analysis in Subsection II-D. Along with the first-order of Taylor expansion $e^{-x}=1-x$, we obtain that when $\bar{\gamma}<2e^{R}-2$, $G_\eta>1$ holds,  and vice versa. $\hfill\blacksquare$
								
								\vspace{20pt}
								Proposition 2 implies that with the higher average SNR, the optimal number of connections is smaller and converges gradually at high SNR. This is due to that the multi-connectivity scheme is more effective in improving reliability when the average SNR is low.
								While in the case of high average SNR, the multi-connectivity scheme consumes more energy, but has the limited PAoI performance gain, resulting in a low EE-PAoI ratio.
								
								\vspace{-6pt}
								\subsection{Risk-Aware, AoI-Optimal and Energy-Efficient Connectivity Scheme}
								Since $\mathcal{P}$1  is an integer programming problem, 
								we first relax the integer constraint in ($C3$) into continuous
								space, and then the optimal number of connections are obtained by rounding. Then we analyze the property of the objective function, as shown in Proposition 3.

								\textbf{Proposition 3} When the number of connections is larger than 2, maximizing the formulated EE-PAoI ratio can be regarded as a CLFP with respect to the number of connections.
								
								\textit{Proof:} {
									The formulated EE-PAoI ratio can be expressed as $\eta=\frac{v(K)L}{g(K)MP_t}$, where $v(K)=\frac{(1-\bar{\epsilon}_K)}{K}$ and $g(K)=\bar{A}_K$.
									In the following, we prove that $v(K)$ is regarded as a concave function and $g(K)$ is a linear function to ensure that maximizing EE-PAoI ratio is a CLFP.
									 By relaxing the number of connections into the continuous space, the second-order derivative of $v(K)$ is given by $\frac{\partial^2 v(K)}{\partial K^2}=-\frac{\partial^2\bar{\epsilon}_K}{\partial K^2}K^3+2K^2\frac{\partial\bar{\epsilon}_K}{\partial K}+2K-2K\bar{\epsilon}_K$.
								Based on (5), the second-order derivative of the average BLEP  in relation to the number of connections is given by $\frac{\partial^2\bar{\epsilon}_K}{\partial K^2}=\int_{0}^{\infty}\Upsilon(\gamma,K)e^{-\frac{\gamma}{\bar{\gamma}}}\epsilon(\gamma)d\gamma$, where $\Upsilon(\gamma,K)=\frac{\gamma^{K-1}\left((\ln(\frac{\bar{\gamma}}{\gamma})+\psi_0(K))^2-\psi_1(K)\right)}{\bar{\gamma}^K\Gamma(K)}$.
								Also, $\psi_0(K)=-0.578+\sum_{n=1}^{K-1}\frac{1}{n}$ and $\psi_1(K)=\sum_{n=0}^{\infty}\frac{1}{(n+K)^2}$ are the first and second derivatives of the Gamma function with respect to the number of connections, respectively.  
								For the case of $K>2$, $\psi_0(K)>0$ and the gap between $\psi_0^2(K)$ and $\psi_1(K)$ is sufficiently small.
								Also, with a large number of connections, $\gamma(K)=(K-1)!$ is large and hence $\Upsilon(\gamma,K)$ results in a small value. By adopting the similar method, the fisrt and second derivatives of BLEP can be regarded as negligible compared to the term of $2K-2K\bar{\epsilon}_K$, resulting in $\frac{\partial^2\bar{\epsilon}_K}{\partial K^2}>0$. The monotonicity of the average PAoI with respect to the number of connections has been proved in Corollary 1, which completes the proof. 
								$\hfill\blacksquare$


								
									
									Based on Proposition 3, $\mathcal{P} 1$ can be solved by using Dinkelbach's algorithm\cite{9690057,Dinkel}.
									Then  the optimal number of connections that balances PAoI performance and EE, namely the risk-aware, AoI-optimal and energy-efficient connectivity scheme, is provided in Lemma 1.
									
									\textbf{Lemma 1} The optimal number of connections that maximizes the EE-PAoI ratio while avoiding risky states is given by $K_\textrm{opt}=\underset{K\in(1,2,K^*)}{\arg\max} \eta$, where 
									\begin{equation}
										K^*=\underset{{K\in\left[K_{\min}, K_{\max}\right]}}{\arg}\max\left\{H(\lambda)\right\},\label{optimal_blocklength}
									\end{equation}		
							can be solved efficiently by the interior point algorithm \cite{Inter} with $H(\lambda)=v(K)-\lambda g(K)$ and $\lambda=\frac{v(K^*)}{g(K^*)}$.  	 	
									
									\textit{Proof}:										It is concluded from ($C1$) that the maximum allowable number of connections is given by $K_{\max}=\frac{P_{\max}}{P_\textrm{t}}$. Also, the minimum allowable number of connections can be obtained as $K_{\min}=\max\{3,\bar{\epsilon}_k^{-1}(1-\frac{\ln{\rm{Pr}}_{\max}}{\lambda(2M-\zeta)}) \}$ based on ($C2$) and the analyzed CLFP condition in Proposition 3.
									Then,  $\mathcal{P}$1 can be viewed as a CLFP in its feasible region. Hence, the solution of $\mathcal{P}$1, \emph{i.e.}, $K^{*}$,  can be obtained by 
									using Dinkelbach’s algorithm\cite{Dinkel} in $\left[K_{\min},K_{\max}\right]$. Then we compare $K^{*}$ with the case of $K=1$ and $K=2$ in terms of EE-PAoI ratio, which yields the optimal number of connections in Lemma 1. 	$\hfill\blacksquare$

									Note that the case of  $K_\textrm{opt}=1$ is also included in the proposed  connectivity scheme.
									Proposition 2 and Lemma 1 indicate that the optimal number of connections is SNR-dependent and affected by the maximum allowable transmit power and violation probability.										
									With a larger maximum allowable transmit power and violation probability, a wider feasible range for the number of connections is obtained.				
									Then, a larger transmit power leads to a lower optimal number of connections.					
									
									\section{Simulation Results}
									In this section, we validate  the derived average AoI/PAoI and PAoI violation probability via Monte Carlo  simulations, where 										100,000 packets are generated with Poisson distribution. We also evaluate  the proposed risk-aware, AoI-optimal and energy-efficient connectivity scheme in WNCSs.
									In this paper, we consider the load frequency control system in smart grid
									as a case study for the WNCS, where the frequency deviation is monitored and controlled to be minimal\cite{LFC}.
									The grid may be unstable if the frequency
									deviation violates a given threshold.
									The parameter setting is given in Table II\cite{LFC,9324753}. 				
											\renewcommand{\arraystretch}{1.3}
									\begin{table}[h]
										\centering
										\caption{Parameter setting}
										\begin{tabular}{lllll}
											\cline{1-2}
											Parameter                                & Value &  &  &  \\ 	\cline{1-2}
											System matrix               & $\mathbf{A}=
											\left[ {\begin{array}{ccc}
													1.17 & 0.67\\
													0.67&0.37 \\
											\end{array} } \right]$  &  &  &  \\ 	\cline{1-2}
											Control matrix  & $\mathbf{B}=\left[0.67, 0.37\right]^{\top}$     &  &  &  \\ \cline{1-2}
											
											
											Process noise variance                   & $\mathbf{R_w}=1e-6\mathbf{I}$     &  &  &  \\ 	\cline{1-2}
											Symbol duration                          & $T_s=0.005$ ms    &  &  &  \\ 	\cline{1-2}
											Number of packets                        & $100000$     &  &  &  \\ 	\cline{1-2}
											
											Maximum  transmit power                   & $P_{\max}=50$ dBm     &  &  &  \\ 	\cline{1-2}
											
											Maximum  violation probability                  & $\rm{Pr}_{\max}=0.1$\%    &  &  &  \\ 	\cline{1-2}
											 PAoI threshold                  &  $\zeta=8$ ms     &  &  &  \\ 	\cline{1-2}
											Arrival rate                        &  $\lambda=1$ packets/ms    &  &  &  \\ 	\cline{1-2}
											Information bits and blocklength          & $L=160$, $m=100$ ch.use    &  &  &  \\ 	\cline{1-2}
										\end{tabular}
									\end{table}
									\subsection{AoI Distribution}

									In Fig. \ref{fig_2}, the impact of the transmit power of each  {connection} on the  average AoI and PAoI with different number of connections is shown.  As expected, as the number of connections and transmit power increase, the average AoI and PAoI are monotonically decreasing, thanks to the higher transmission reliability.  Also, it is shown that the reduction of the average AoI/PAoI is significant with low transmit power and small number of connections. This is due to that with a larger number of connections and higher transmit power, the reliability gain is limited and hence the average AoI/PAoI reduction is insignificant.
									Furthermore, it is shown that the average PAoI and AoI have similar trends. Hence, we focus on the PAoI performance in the following simulations.
									
									\begin{figure}[htbp]
										\centering
										\includegraphics[width=7.8cm]{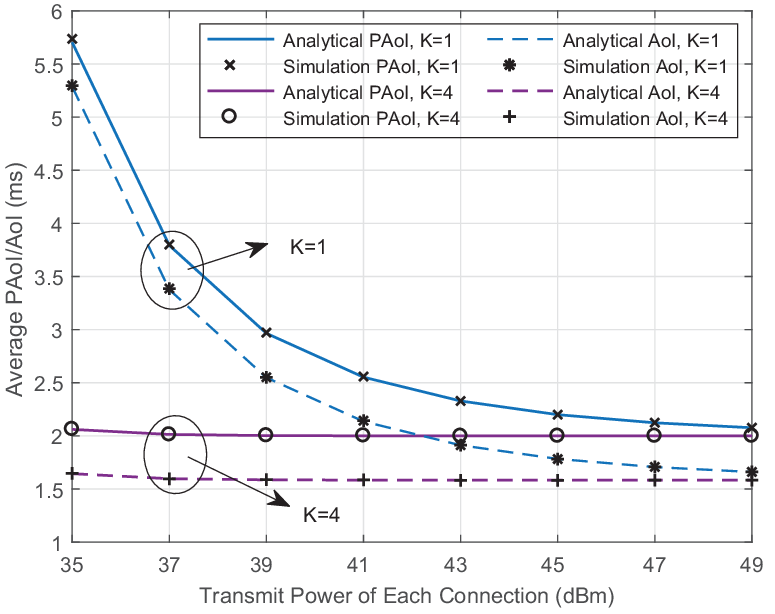}
										\caption{The average AoI/PAoI vs.  transmit power of each {connection} with different number of connections.}\label{fig_2}
									\end{figure}
							
									%


									%

									\begin{figure}[htbp]
										\centering
										\begin{minipage}[t]{0.48\textwidth}
											\centering
											\includegraphics[width=8.5cm]{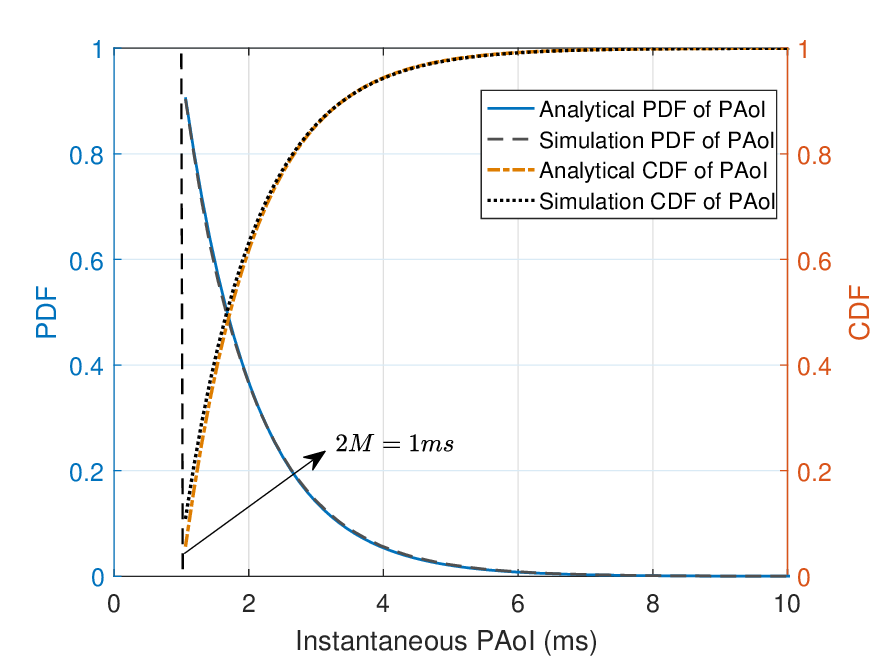}
											\caption{The PDF and CDF of PAoI  with the transmit power $P_\textrm{t}=35$ dBm and $K=4$  {connections}.}\label{fig_6}
										\end{minipage}
										\hspace{8pt}
										\begin{minipage}[t]{0.48\textwidth}
											\centering
											\includegraphics[width=8.5cm]{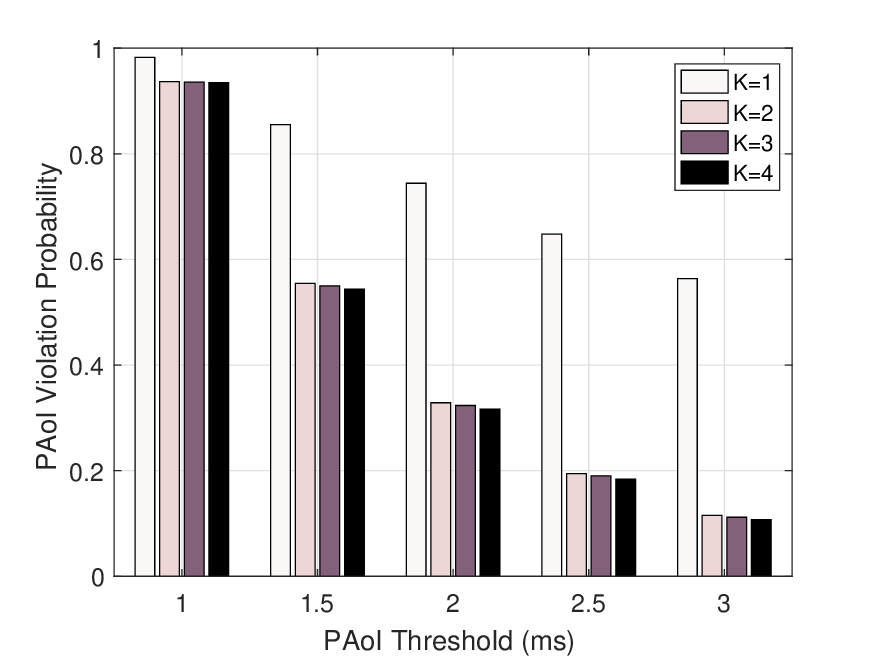}
											\caption{Impact of the number of connections on the PAoI violation probability  with the transmit power $P_\textrm{t}=35$ dBm and different PAoI thresholds.}\label{fig_7}
										\end{minipage}
									\end{figure}
									In Fig. \ref{fig_6}, the PDF and CDF of PAoI are presented, with the transmit power $P_\textrm{t}=35$ dBm and the number of connections $K=4$.
									It is shown that the derived PDF and CDF of PAoI in (14) and (15) match the simulation results accurately. 
								 	As stated in Theorem 4, the PDF of PAoI is zero when the instantaneous PAoI is smaller than $2M$, and is an exponential function of PAoI when the instantaneous PAoI is larger than $2M$. This demonstrates that  the PAoI violation probability is a negative exponential function of PAoI threshold. 
									
									Fig. \ref{fig_7} illustrates the impact of PAoI threshold and number of connections on the PAoI violation probability. It is shown  that the PAoI violation probability can be significantly reduced  by using the multi-connectivity scheme, especially with a large PAoI threshold and small number of connections.
									This is due to that with a larger PAoI threshold, the PAoI violation probability is more sensitive to the average BLEP, as shown in (16).
									This also verifies that increasing the number of connections is not always effective, as the reliability gain from multiple connections is limited.
									Specifically, Fig. 5 shows that the PAoI violation probability with a  dual connection  can be reduced by up to 5-fold with the PAoI threshold of $3$ ms, compared to the single-connectivity scheme.
									
									%

									
									\subsection{Age-Energy Tradeoff}

									Fig. \ref{fig_8} shows the impact of the number of connections on the average PAoI and total energy consumption with different transmit powers, indicating the tradeoff between PAoI and energy.
									This motivates us to find the optimal number of connections that ensures both PAoI and EE performance.
									It is concluded that with  low transmit power, adopting the multi-connectivity scheme is able to reduce the average  PAoI significantly.
									Also, it is shown that the average PAoI  decreases monotonically as the number of connections increases, and gradually converges at a large number of connections. This is due to that with a large number of connections, there is a limited improvement in reliability as the number of connections increases, leading to a  limited reduction in PAoI violation probability.
									In particular, with the arrival rate $\lambda=1$ packets/ms, block length $m=100$ c.u. and symbol duration $T_\textrm{s}=0.005$ ms, the minimum average PAoI is $\frac{1}{\lambda}+2M=2$ ms, as provided in Corollary 1.

										

												\begin{figure}[htbp]
													\centering
													\begin{minipage}[t]{0.48\textwidth}
														\centering
														\includegraphics[width=7.5cm]{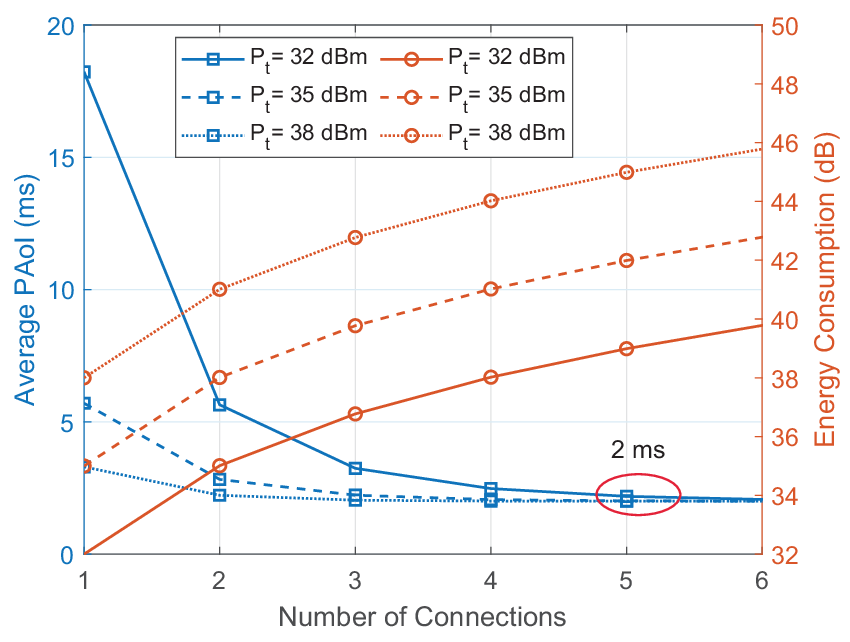}
														\caption{Average PAoI vs. number of connections with different transmit power.}\label{fig_8}
													\end{minipage}
													\hspace{8pt}
													\begin{minipage}[t]{0.48\textwidth}
														\centering
														\includegraphics[width=7.8cm]{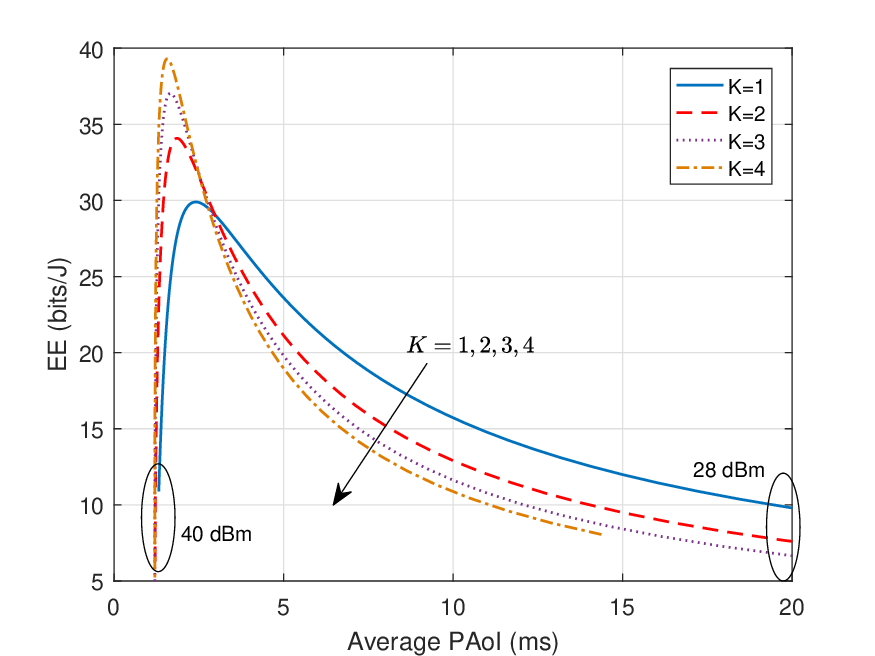}
														\caption{PAoI-EE tradeoff with different number of connections.}\label{fig_9}
													\end{minipage}
												\end{figure}

												In Fig. \ref{fig_9}, the tradeoff between the average PAoI and EE is shown, where the coding rate is fixed and the transmit power varies from 28 dBm to 40 dBm. As the transmit power increases, the average PAoI is monotonically decreasing,  while the EE increases  at first and then decreases with high transmit powers. This is due to that, the reliability gain is limited and the energy consumption is high with high transmit powers, resulting in low EE. 
												Also, the results show that with a given EE, the multi-connectivity scheme achieves a lower average PAoI, indicating the better AoI-energy tradeoff performance.

												%
												%
												
												\subsection{Risk-Aware, AoI-Optimal and Energy-Efficient Connectivity Scheme}
												
												In Fig. \ref{fig_10}, the impact of the number of connections on the EE-PAoI ratio is shown. It is shown that there exists an optimal number of connections that maximizes the EE-PAoI ratio, as proved in Proposition 2.
												This is due to that with a large number of connections, the  EE is low and the increment of the PAoI performance gain is limited. 
												Also, the optimal number of connections is lower when the transmit power is higher, thanks to the higher transmission reliability, as stated in Proposition 2.
												Then we consider the load frequency control system in smart grid as a case study for the WNCS.  
										
										Fig. \ref{fig_11} shows the plant state (\emph{i.e.}, frequency deviation)  with the optimized number of connections under transmit power $P_\textrm{t}=35$ dBm, where the PAoI violation probability is limited to less 0.1\%  with the PAoI threshold of 8 ms. It is shown that with the optimized number of connections, the state violation probability is 0.047\%. This demonstrates the effectiveness of our proposed risk-aware connectivity scheme, which can avoid risky plant states by limiting the  PAoI violation probability.											
												
												\begin{figure}[tbp]
													\centering
													\begin{minipage}[t]{0.48\textwidth}
														\centering
														\includegraphics[width=7.5cm]{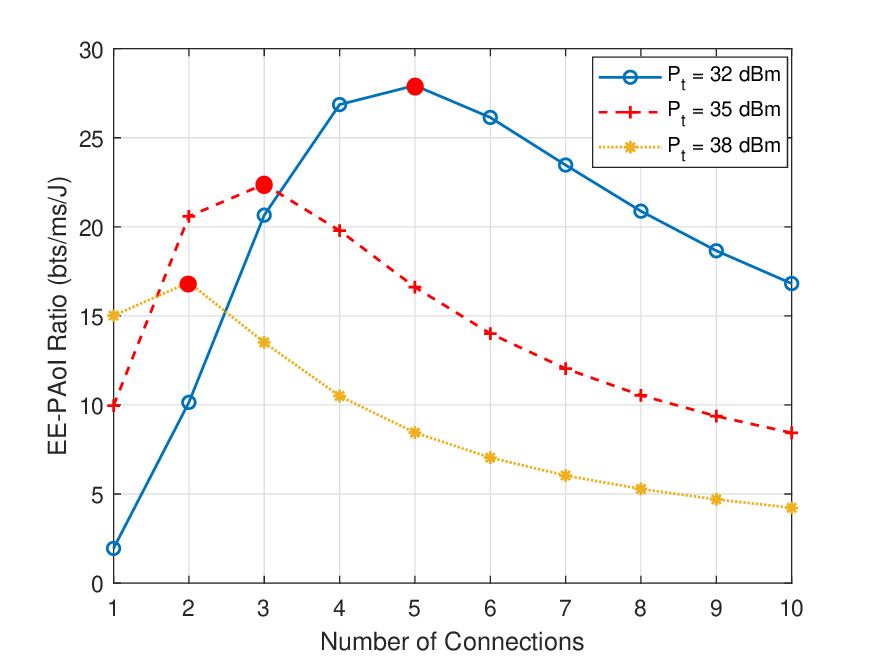}
														\caption{EE-PAoI ratio vs. number of connections with different transmit power.}\label{fig_10}
													\end{minipage}
													\hspace{8pt}
													\begin{minipage}[t]{0.48\textwidth}
														\centering
														\includegraphics[width=7.5cm]{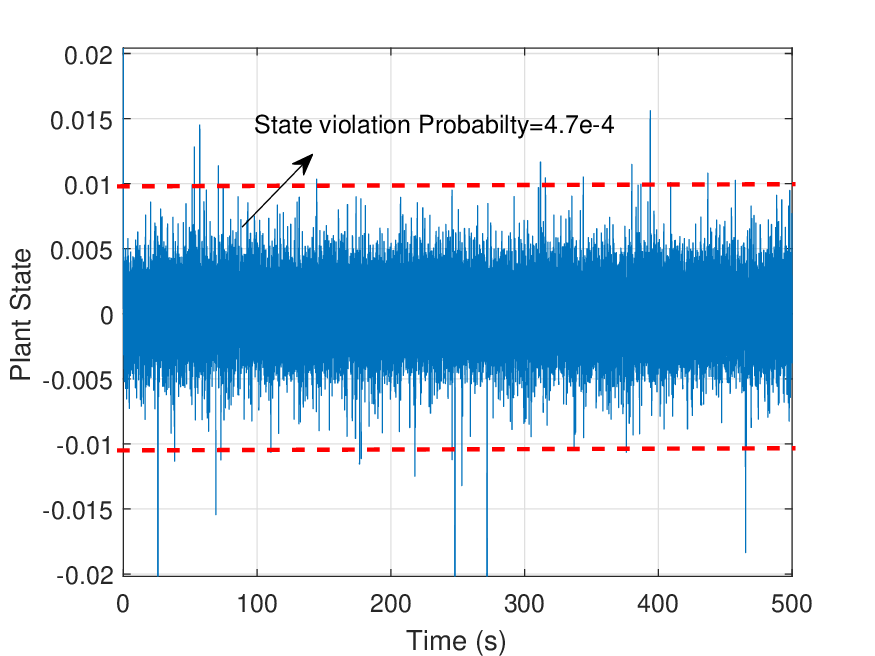}
														\caption{Plant state in the WNCS (\emph{i.e.}, frequency deviation in smart grid) with the optimized number of connections and transmit power $P_\textrm{t}=35$ dBm.}\label{fig_11}
													\end{minipage}
												\end{figure}


													Fig. \ref{fig_13} shows that  the proposed AoI-optimal and energy-efficient connectivity scheme achieves a higher EE-PAoI ratio, especially with the low transmit powers.
													For example, with the transmit power of $P_\textrm{t}=32$ dBm, the proposed connectivity scheme achieves a 15-fold improvement of EE-PAoI ratio, compared to the single-connectivity scheme.

											Fig. \ref{fig_14} shows the impact of transmit power on the optimal number of connections. It is concluded that the optimal number of connections is decreasing with respect to the transmit power. Also, the derived threshold that depends on the coding rate,  characterizes the superior region of the energy-efficient scheme for AoI-oriented systems accurately, as demonstrated in Proposition 2.
													
																\begin{figure}[htbp]
														\centering
														\begin{minipage}[t]{0.48\textwidth}
															\centering
															\includegraphics[width=7.5cm]{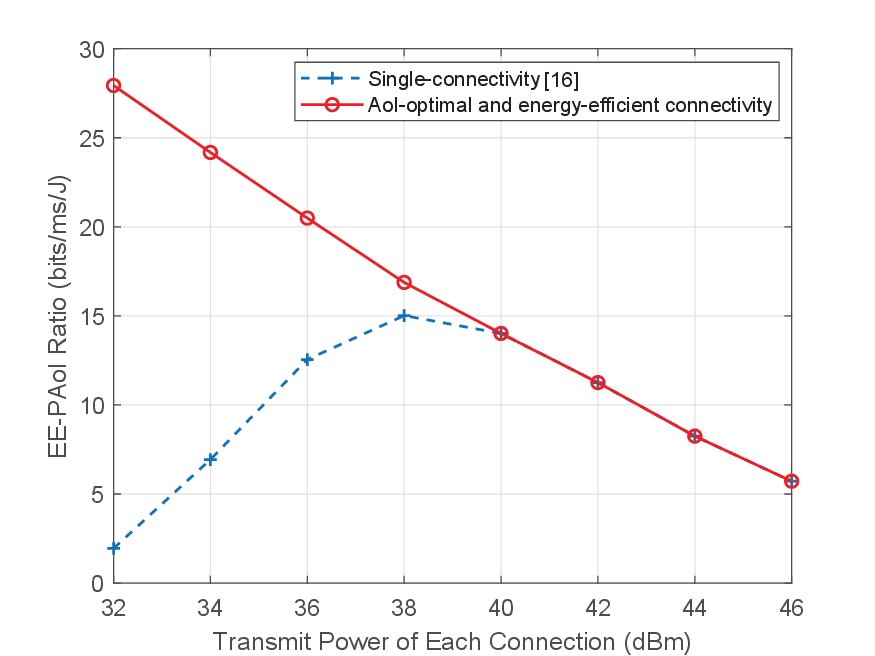}
															\caption{EE-PAoI ratio vs. transmit power of each connection.}\label{fig_13}
														\end{minipage}
														\hspace{8pt}
														\begin{minipage}[t]{0.48\textwidth}
															\centering
															\includegraphics[width=7.5cm]{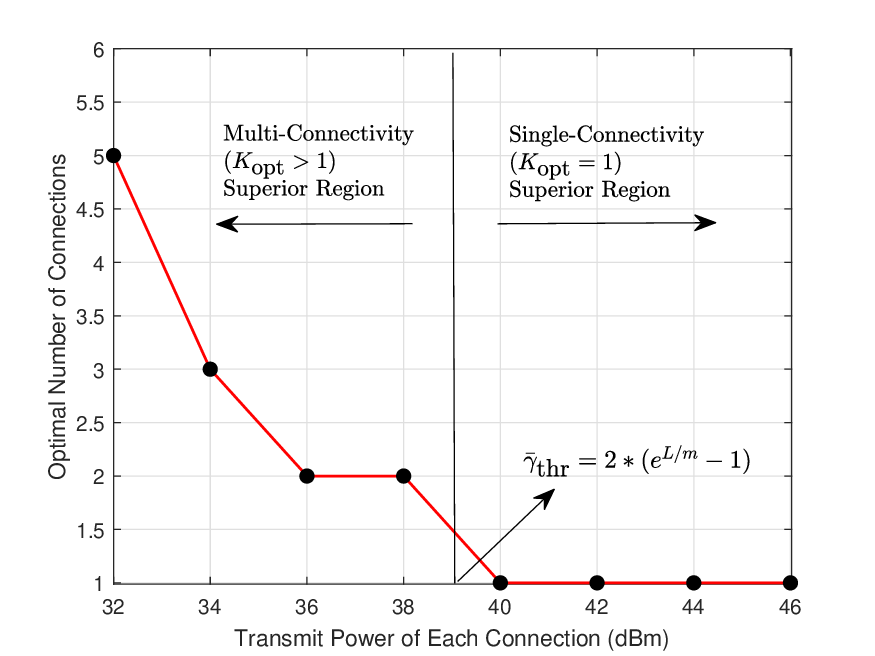}
															\caption{Optimal number of connections vs. transmit power of each connection.}\label{fig_14}
														\end{minipage}
													\end{figure}

															
															\section{Conclusion}
															
															In this paper, we have investigated the AoI and EE performances of the  multi-connectivity short-packet WNCS with transmission diversity. We have conducted a comprehensive analysis of both the average AoI/PAoI  and the distribution of PAoI for timely transmission and reliable control.
															To address the trade-off between risk, freshness and energy efficiency,
															we have proposed a risk-aware, AoI-optimal and energy-efficient connectivity scheme to find the optimal number of connections that maximizes the EE-PAoI ratio while avoiding risky states.
															Particularly, we have proved  that the multi-connectivity scheme can achieve a higher EE-PAoI with a low average SNR, compared to the single-connectivity scheme. In particular, the SNR threshold for connectivity mode switching has been derived in closed form. 
															Simulation results have demonstrated that the proposed connectivity scheme is able to avoid risky states and achieve a 15-fold improvement of EE-PAoI ratio at low average SNR, compared with the single-connectivity system.
In the future, we will consider a more
practical WNCS with two-way imperfect links and investigate the joint optimization of coding rate, power allocation and link selection.

\section*{Acknowledgment}
This work is a part of project “Future Proof Reliable and Resilient Wireless
Communications for Virtual Power Plant (W-VPP),” supported by National
Research Foundation (NRF) Singapore, under its Industry Alignment Fund
(Pre-Positioning) (IAF-PP) for Urban Solutions and Sustainability (USS)
Domain, Research Innovation Enterprise 2020 Plan (RIE2020), and Energy
Grid 2.0 Programme; and in part by the National Natural Science Foundation of
China under Grants 62171161.

															\appendices


															\section{Proof of Theorem 1}
															
															Fig. 2 indicates that the inter-departure time $Y^{\text{NR}}$ is composed of the service time $S^{\text{NR}}$ and the waiting time $W^{\text{NR}}$. 
Since the arriving packets follow  the Poisson distribution with the average rate $\lambda$, 
 the expectation of the waiting time $W^{\text{NR}}$ is provided as $\mathbb{E}[W^{\text{NR}}]=\frac{1}{\lambda}$\cite{wangrui}.
By considering the average BLEP $	\bar{\epsilon}_\textrm{K}$, the expectation of the inter-departure time is given by 
															\begin{equation}
																\mathbb{E}[Y^{\text{NR}}]=(1-	\bar{\epsilon}_\textrm{K})M+	\bar{\epsilon}_\textrm{K}\mathbb{E}[M+W^{\text{NR}}+Y^{\text{NR}}],\tag{17}
															\end{equation} which is obtained by  using a recursive method and can be expressed as
															\begin{equation}
																\mathbb{E}[Y^{\text{NR}}]=\frac{\frac{1}{\lambda}+M}{1-	\bar{\epsilon}_\textrm{K}}.\tag{18}
															\end{equation} Substituting $\mathbb{E}[Y^{\text{NR}}]$ and $S^{\text{NR}}=M$ into (2) yields the average PAoI in (8).
														Based on the independence among $W^{\text{NR}}$, $M$ and $Y^{\text{NR}}$, we have \cite{wangrui}
															\begin{equation}
																\mathbb{E}[{Y^{\text{NR}}}^2]=\frac{(M+\frac{1}{\lambda})^2(1+	\bar{\epsilon}_\textrm{K})}{(1-	\bar{\epsilon}_\textrm{K})^2}+\frac{1}{\lambda^2(1-	\bar{\epsilon}_\textrm{K})}.\tag{20}
															\end{equation}  
															Then, substituting $\mathbb{E}[{Y^{\text{NR}}}]$ and  $\mathbb{E}[{Y^{\text{NR}}}^2]$ into  (3) yields the average AoI in (9).

												\section{Proof of Theorem 2}
												Based on the assumption of ARQ retransmission, the expectation of service time can be obtained as $\mathbb{E}[S^{\text{ARQ}}]=\frac{M}{1-\bar{\epsilon}}$.
											Similar to the analysis in Appendix A, the expectation of waiting time is given by  
												 $\mathbb{E}[W^{\text{ARQ}}]=\frac{1}{\lambda}$\cite{wangrui}. By considering the average BLEP $	\bar{\epsilon}$ and ARQ retransmission, the average inter-depature time is obtained as
												\begin{equation}
													\mathbb{E}[Y^{\text{ARQ}}]=(1-\bar{\epsilon})({1}/{\lambda}+M)+\bar{\epsilon}(M+	\mathbb{E}[Y^{\text{ARQ}}])
												\end{equation}
												which leads to $\mathbb{E}[Y^{\text{ARQ}}]=\frac{1}{\lambda}+\frac{M}{1-\bar{\epsilon}}$.
												Substituting $\mathbb{E}[Y^{\text{ARQ}}]$ and $\mathbb{E}[S^{\text{ARQ}}]$ into (2) yields the average PAoI in (10).
												Then we have 
												\begin{small}
												\begin{equation}
													\begin{aligned}
														\mathbb{E}[{Y^{\text{ARQ}}}^2]=&(1-\bar{\epsilon})\left(\frac{2}{\lambda^2}+M^2+\frac{2M}{\lambda}\right)\\
														&+\bar{\epsilon}\left(M^2+	\mathbb{E}[{Y^{\text{ARQ}}}^2]+2M\mathbb{E}[Y^{\text{ARQ}}]\right),	
													\end{aligned}\tag{18}
												\end{equation} 
												\end{small}which can be further simplified as 
												\begin{equation}
													\begin{aligned}
														\mathbb{E}[{Y^{\text{ARQ}}}^2]=&\frac{2(1-\bar{\epsilon}_K)+\lambda^2+M^2+2\lambda M(1-\epsilon)}{\lambda^2}\\&+\frac{2M\bar{\epsilon}_K(1-\bar{\epsilon}_K+\lambda M)}{\lambda(1-\bar{\epsilon}_K)}.
													\end{aligned}\tag{19}
												\end{equation}  
												Then, substituting $\mathbb{E}[{Y^{\text{ARQ}}}]$ and  $\mathbb{E}[{Y^{\text{ARQ}}}^2]$ into  (3) yields the average  AoI in (11).

														\end{document}